\documentclass[%
 reprint,
showpacs,preprintnumbers,
 amsmath,amssymb,
 aps,
]{revtex4-1}

\usepackage{graphicx}

\newcommand{\myfig}[4][ht]{
\begin{figure}[#1]
\centering
\includegraphics[#2]{#3}
\caption{#4\label{#3}}
\end{figure}
}

\newcommand{\myfigwide}[4][ht]{
\begin{figure*}[#1]
\centering
\includegraphics[#2]{#3}
\caption{#4\label{#3}}
\end{figure*}
}

\newcommand{\revision}[1]{%
{#1}%
}

\usepackage{dcolumn}
\usepackage{bm}

\begin{document}


\title{SUPERDIFFUSIVE HEAT CONDUCTION IN SEMICONDUCTOR ALLOYS\\[2mm]II. Truncated L\'evy formalism for experimental analysis}

\author{Bjorn Vermeersch$^1$}%
 \email{bvermeer@purdue.edu}

\author{Amr M.S. Mohammed$^1$}%

\author{Gilles Pernot$^2$}

\author{Yee Rui Koh$^1$}%

\author{Ali Shakouri$^1$}%
 \email{shakouri@purdue.edu}
\affiliation{$^1$ Birck Nanotechnology Center, Purdue University, West Lafayette, IN 47907, USA \linebreak[4] $^2$ LOMA, Universit\'e de Bordeaux 1, Talence, 33400, France}

\date{\today}

\begin{abstract}
Nearly all experimental observations of quasi-ballistic heat flow are interpreted using Fourier theory with modified thermal conductivity. Detailed Boltzmann transport equation (BTE) analysis, however, reveals that the quasi-ballistic motion of thermal energy in semiconductor alloys is no longer Brownian but instead exhibits L\'evy dynamics with fractal dimension $\alpha < 2$. Here, we present a framework that enables full 3D experimental analysis by retaining all essential physics of the quasi-ballistic BTE dynamics phenomenologically. A stochastic process with just two fitting parameters describes the transition from pure L\'evy superdiffusion as short length and time scales to regular Fourier diffusion. The model provides accurate fits to time domain thermoreflectance raw experimental data over the full modulation frequency range without requiring any `effective' thermal parameters \revision{and without any a priori knowledge of microscopic phonon scattering mechanisms}. Identified $\alpha$ values for InGaAs and SiGe match ab initio BTE predictions within a few percent. Our results provide experimental evidence of fractal L\'evy heat conduction in semiconductor alloys. The formalism additionally indicates that the transient temperature inside the material differs significantly from Fourier theory and can lead to improved thermal characterization of nanoscale devices and material interfaces.
\end{abstract}

\maketitle

\subsection*{Introduction}
Heat in non-metallic solids is predominantly conducted by random motion of energy carriers called phonons \cite{phonons}. The spectrum of phonon mean free paths (MFPs), i.e. the distribution of the average distance phonons travel ballistically between consecutive scattering events, governs key aspects of the thermal behavior. Even at room temperature, a significant portion of heat in commonly used semiconductors is found to be carried by phonons with MFPs well into the micron range \cite{MFP1,MFP2}. Thermal transport over these length scales is of crucial importance for nanoscale devices \cite{devices}.
\par
The macroscopic net result of the phonon dynamics is observed as redistribution of thermal energy typically described by the Fourier diffusion equation. However, when the dimensions of the thermal gradient become comparable to phonon MFPs, this classical model begins to fail \cite{boltzmann}. Several hyperbolic heat conduction \cite{heatwaves} and ballistic-diffusive \cite{ballisticdiffusive} theories were proposed but many characteristic features they predicted have not been observed experimentally. Measurements in which the physical dimensions \cite{siemens}\nocite{minnich}--\cite{johnson} or penetration depth \cite{cahill,malen} of the heat source overlap with phonon MFPs exhibit an apparent reduction of effective thermal conductivity \cite{minnich}--\cite{malen} $k_{\text{eff}}$, or equivalently, an additional ballistic thermal resistance \cite{siemens}. Despite showing clear evidence of nondiffusive thermal transport, most experiments are interpreted with modified Fourier theory, i.e. a regular diffusion model but with adjusted thermal parameters.
\par
Boltzmann transport equation (BTE) analysis with ab initio phonon dispersions and scattering rates, documented in part I of this paper \cite{part1}, shows that quasi-ballistic transport in semiconductor alloys is no longer Brownian, but instead governed by a L\'evy process with fractal dimension $\alpha < 2$. The associated energy density distribution is non-Gaussian, and a new approach beyond modified Fourier theory is needed to accurately represent the quasi-ballistic transport dynamics. BTE solutions themselves, unfortunately, are not easily suitable for direct comparison with experiments. Analytical BTE modeling is typically limited to 1D analysis of the dominant cross-plane heat flow, leaving the method unable to account for lateral heat spreading and Gaussian shape of the heat source encountered in actual measurements. In addition, non-idealities in real world samples such as crystal impurities or grain boundaries cause the phonon spectra and resulting thermal properties to invariably deviate to some extent from ab initio predictions. 
\par
Here, we provide a phenomenological approach that preserves the essential dynamics contained within BTE solutions yet at the same time offers sufficient flexibility to achieve full 3D analysis of experimental measurements. \revision{We stress that the BTE analysis performed in Part I only serves to provide physical support for a L\'evy-based approach, but is not directly involved in any way in the actual processing of measurement data. The formalism we present below is a fully autonomous model, independent of the relaxation time approximation, in which key properties of L\'evy dynamics act as free fitting parameters. The method is widely applicable to heat conduction dominated by mass impurity (or other high order) phonon scattering, without requiring ab initio simulations or any other prior knowledge about the phonon properties of the specific material at hand.} Our formalism provides superior fittings of raw thermoreflectance (TR) experimental data without the need for any `effective' thermal parameters varying with laser modulation frequency. Identified L\'evy fractal dimensions in InGaAs and SiGe match ab initio BTE predictions within a few percent. In addition, the formalism offers great potential for improved thermal performance characterization of nanoscale devices and metal/semiconductor interfaces.
\subsection*{Methodology}
Our formalism is based on a probabilistic framework: the motion of thermal energy inside the semiconductor is described in terms of a stochastic process. The method relies on the duality $T(x,t) \leftrightarrow P(x,t)/C$ between the temperature response $T$ to a unit energy impulse and probability $P$ to find a random walking energy carrier in position $x$ at time $t$. $C$ denotes the volumetric heat capacity of the medium. Continuous time random walk (CTRW) processes essentially consist of a series of transition events \cite{CTRWmaster}. Each transition increments the position of the energy carrier by an amount $u$ randomly chosen from a distribution $p_U(u)$, while the time between consecutive transitions is governed by a distribution $p_T(t)$. In our context, the process is Poissonian in time, $p_T(t) = \Theta \exp(-\Theta t)$ with $\Theta$ the average number of transitions per second, while $p_U$ is always an even function. The latter expresses equal probability for left and right transitions in accordance with thermal isotropy.
\par
The mathematical description of the process simplifies considerably when $p_U$ and $p_T$ are stochastically independent. Physically, this decoupling of space and time results in a situation in which the  transition velocity can sporadically become arbitrarily large. Although this is not a rigorous representation of actual phonon dynamics, an unbounded velocity approximation is physically adequate for our purposes. Justification follows by comparing the extent of the thermal gradient $\Delta x \simeq \ell$ to the energy containment $|x| \leq x_{\text{max}}$ imposed by finite phonon velocities. Here, $\ell = \sqrt{2Dt}$ is the Fourier thermal penetration depth with $D = \kappa/C$ the thermal diffusivity of the medium, and $x_{\text{max}} = vt$ with $v$ the sound velocity. Time domain TR observations utilise data taken at pump-probe delays $t_{\text{pp}} \geq 50\,$ps for thermal characterization. Even at such short times, we find that $x_{\text{max}}/\ell > 10$ in typical semiconductor alloys ($D \approx 5\,$mm$^2$/s, $v\approx 5000\,$m/s). The finite velocity thus hardly imposes restrictions on the development of the thermal gradient at time scales probed by the experiments, and can be safely considered as effectively unbounded. The transitions in the resulting stochastic process with uncoupled space and time are typically called `flights' with `jump lengths' governed by $p_U$.
\par
Regular Fourier diffusion is stochastically equivalent to Brownian motion \cite{brownian}. This process obeys a jump length probability distribution $p_{\text{U}}(u) \propto |u|^{-3}$ and induces the familiar Gaussian energy density with mean square displacement (MSD) $\ell^2(t) = 2Dt$. During the quasi-ballistic regime, however, BTE solutions of the energy density are governed by a L\'evy process with fractal dimension $\alpha$ between 1 and 2 \cite{part1}. These dynamics correspond to jump length distributions $p_{\text{U}}(u) \propto |u|^{-(1+\alpha)}$ and induce characteristic fractal patterns \cite{fractalwalks} consisting of medium range motion clusters separated by occasional long jumps (Fig. 1a). L\'evy behavior has been observed in travel patterns of foraging animals \cite{animals}, protein movements along DNA chains \cite{DNA}, turbulence in fluids \cite{fluidturbulence}, and financial market fluctuations \cite{stockmarket}. Similar effects were observed in the context of anomalous heat conduction in theoretical studies of 1D atomic chains between reservoirs at constant temperature \cite{atomchain}. The latter is still quite different from realistic experimental and technological configurations. The model we develop here describes 3D quasi-ballistic heat flow in thermally semi-infinite structures subjected to a transient surface heat flux.
\myfigwide[!htb]{width=0.8\textwidth}{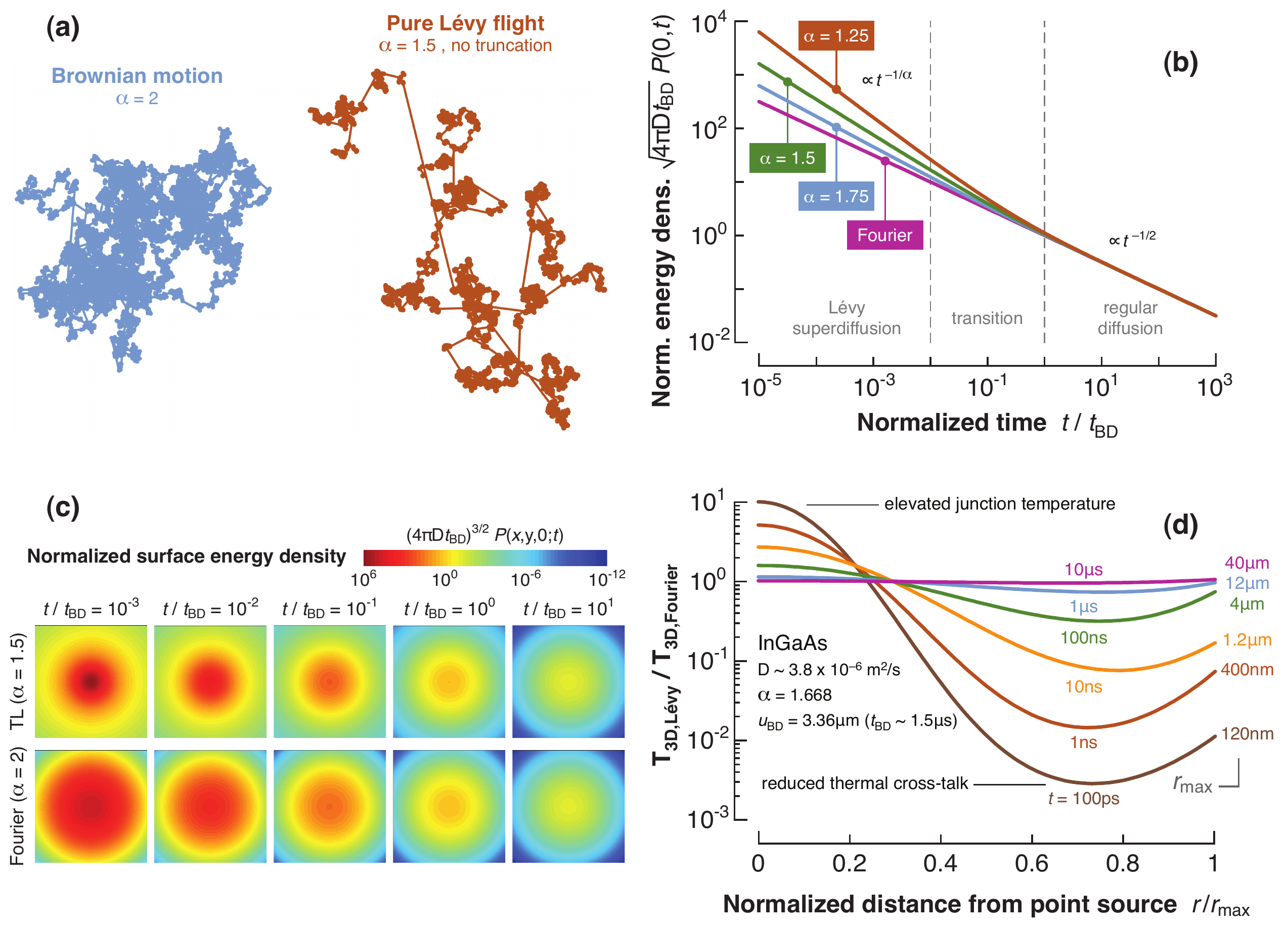}{\textbf{Stochastic energy transport in solid media.} (a), Exemplary random walk trajectories with 10,000 steps each. (b), Normalized source response for 1D truncated L\'evy heat flow in infinite medium. The regime transition from pure L\'evy superdiffusion to regular Fourier diffusion is clearly visible. (c), Dimensionless energy pulse response for 3D heat flow in semi-infinite medium. Each square shows a thermal map over a region measuring $10 \ell(t) \times 10 \ell(t)$ centered around the point heat source, with $\ell(t) = \sqrt{2Dt}$ the Fourier thermal penetration length. (d), Ratio of truncated L\'evy and Fourier 3D single pulse responses on semi-infinite InGaAs at room temperature. Listed parameters were determined from TDTR experiments presented below.}
\par
The inherent fractal nature of pure L\'evy flights maintains the quasi-ballistic regime indefinitely. In reality, however, the thermal transport recovers to regular Fourier diffusion at sufficiently long length and time scales. This gradual transition between the two regimes is naturally embedded in the BTE framework. The same can be achieved here by considering a truncated L\'evy (TL) process, in which the likelihood of very long jumps is suppressed \cite{truncation1,truncation2}. The main principle is to force the tail of $p_U$ to drop equally or more steeply than $|u|^{-3}$ such that Brownian dynamics are recovered at sufficiently large distances. For mathematical convenience, we use exponential truncation of the form
\begin{equation}
p_{\text{U}}(u) = \frac{A \exp \left( - |u|/u_{\text{BD}} \right)}{|u|^{(1+\alpha)}} \label{jumplength}
\end{equation}
with $A$ a normalizing constant determined below and $\alpha , u_{\text{BD}}$ two fitting parameters. As demonstrated later, $u_{\text{BD}}$ and associated $t_{\text{BD}} = u_{\text{BD}}^2/2D$ regulate respectively the length and time scales over which the transition from quasi-ballistic to diffusive transport occurs. We note here that certain subtypes of exponentially truncated L\'evy flight processes can be described in terms of generalized fractional diffusion equations \cite{TLfractional}. This may offer prospects for future development of a universal heat equation for microscale thermal transport in closed form.
\par
The generalized CTRW master equation \cite{CTRWmaster} provides the single pulse energy density in Fourier-Laplace domain, $\mathcal{P}(\xi,s)$, for given jump length and wait time distributions in transformed variables $p_U(\xi)$ and $p_T(s)$. For a Poissonian process with jump frequency $\Theta$ and stochastically independent jump length distribution $p_U(u)$, the solution can be inverted analytically to the time domain, and we obtain:
\begin{equation}
\ln \mathcal{P} (\xi,t) = -\Theta t \int \limits_{-\infty}^{\infty} \left[ 1 - \exp(-j \xi u ) \right] p_{\text{U}}(u) \mathrm{d} u \label{lnPgeneral}
\end{equation}
in which $j$ denotes the complex unit. For the truncated L\'evy jump length distribution described by (\ref{jumplength}) this can be evaluated analytically:
\begin{multline}
1 < \alpha < 2 : \quad \ln \mathcal{P} (\xi,t) = (-2t) \frac{\pi A \Theta}{\Gamma(\alpha) \alpha \sin(\alpha \pi)} \times  \label{lnP} \\ \left[ (\xi^2 + \xi_{\text{BD}}^2)^{\alpha/2} \cos \left( \alpha \arctan \left[ \frac{\xi_{\text{BD}}}{|\xi|} \right] - \frac{\alpha \pi}{2}\right) - \xi_{\text{BD}}^{\alpha} \right]
\end{multline}
with $\xi_{\text{BD}} = u_{\text{BD}}^{-1}$. The energy density in real space-time domain then directly follows from numerical inverse Fourier transform:
\begin{equation}
P_{\text{1D}}(x,t) = \frac{1}{\pi} \int \limits_{0}^{\infty} \mathcal{P}(\xi,t) \cos(\xi x) \mathrm{d}\xi \label{P1D}
\end{equation}
where we used that $\mathcal{P}(\xi,t)$ is even in $\xi$. At long times $t \gg t_{\text{BD}}$, $\mathcal{P}$ is strongly exponentially dampened, and the integral (\ref{P1D}) is completely dominated by small wavenumbers $\xi \ll \xi_{\text{BD}}$. Series expansion of (\ref{lnP}) for this regime yields
\begin{equation}
\xi \ll \xi_{\text{BD}} : \quad \ln \mathcal{P}(\xi,t) \rightarrow -\xi^2 t \left[ \frac{\pi A \xi_{\text{BD}}^{\alpha-2} (1-\alpha)}{\Gamma(\alpha) \sin(\alpha \pi)} \right] \label{lnPlimit}
\end{equation}
The proportionality to $\xi^2$ signals recovery of regular Fourier diffusion $\ln \mathcal{P}_{\text{F}} = -D \xi^2 t$. Comparison of the latter to (\ref{lnPlimit}) sets the prefactor in (\ref{lnP}):
\begin{equation}
\frac{\pi A \Theta}{\Gamma(\alpha) \alpha \sin (\alpha \pi)} = \frac{\xi_{\text{BD}}^{2-\alpha}D}{\alpha (1-\alpha)}
\end{equation}
such that we finally have
\begin{multline}
1 < \alpha < 2 : \quad \ln \mathcal{P}(\xi,t) = - \frac{2 \xi_{\text{BD}}^2 D t}{\alpha (1-\alpha)} \times \\ \left[ (\tilde{\xi}^2 + 1)^{\alpha/2} \cos \left( \alpha \, \mathrm{arctan}(|\tilde{\xi}|^{-1}) - \frac{\alpha \pi}{2} \right) - 1 \right] \label{lnPfinal}
\end{multline}
with $\tilde{\xi} = \xi/\xi_{\text{BD}}$. \revision{For the sake of completeness, we note that the solution for special case $\alpha = 1$ does not simply follow by taking the $\alpha \rightarrow 1$ limit of (\ref{lnPfinal}) but instead must be derived directly from (\ref{lnPgeneral}) which eventually yields}
\begin{multline}
\alpha = 1 : \quad \ln \mathcal{P}(\xi,t) = - \xi_{\text{BD}}^2 D t \times \left[ \pi \tilde{\xi} H(\xi - 1) \right. \\
\left. - \ln(1+\tilde{\xi}^2) - \tilde{\xi} \arctan \left( \frac{2 \tilde{\xi}}{\tilde{\xi}^2 - 1} \right) \right] \label{lnPalpha1}
\end{multline}
At short times $t \ll t_{BD}$, the behavior of the TL process is dominated by large spatial frequencies $\tilde{\xi} \gg 1$. According series expansion of (\ref{lnPfinal}) or (\ref{lnPalpha1}) shows this corresponds to a pure L\'evy regime $\ln \mathcal{P}_{\text{L}} = - D_{\alpha} |\xi|^{\alpha} t$ with fractional diffusivity
\begin{eqnarray}
1 < \alpha < 2: \quad D_{\alpha} & = & \frac{2 D \cos \left[ (1-\alpha/2) \pi \right]}{\alpha (\alpha-1) u_{\text{BD}}^{2-\alpha}} \label{Dalpha} \\
\alpha = 1 : \quad D_{\alpha} & = & \frac{\pi D}{u_{\text{BD}}}
\end{eqnarray}
This expression lends a deeper, microscopic meaning to $u_{\text{BD}}$ that goes beyond the intuitive macroscopic notion of characteristic transition length scale.\par%
Once numerical evaluation of $P_{\text{1D}}(x,t)$ is performed, the formalism can be easily extended to 3D heat flows based on isotropy and symmetry arguments. The single pulse temperature response for a point source on a semi-infinite medium becomes
\begin{equation}
T_{\text{3D}}(r,t) = \frac{2}{C} P_{\text{1D}}^3(x=r/\sqrt{3},t)
\end{equation}
with $r$ the distance from the source. From this we then obtain the Green's function $G(h,f)$ of the semiconductor surface for truncated L\'evy transport in the Fourier-Hankel domain:
\begin{equation}
G(h,f) = \int \limits_{0}^{\infty} \int \limits_{0}^{\infty} T_{\text{3D}} (r,t) \exp(-j 2\pi f t) J_0(hr) r \mathrm{d}r \mathrm{d}t
\end{equation}
These operations must be performed numerically since $T_{\text{3D}}(r,t)$ is not known in closed form. A simple quadrature scheme
\begin{eqnarray}
G(h,t) & \approx & \frac{1}{2h} \sum \limits_{n} \left[ T_{\text{3D}}(r_n,t) + T_{\text{3D}}(r_{n+1},t) \right] K(r_n,r_{n+1}) \nonumber \\
K(a,b) & = & b J_1(hb) - a J_1(ha)
\end{eqnarray}
suffices for the Hankel transform. For the subsequent transform to the frequency domain we use a collocation scheme \cite{FFTcollocation} with logarithmic time grid. The resulting $G(h,f)$ then simply replaces the conventional Fourier diffusion kernel $G_{\text{F}}(h,f) = ( 2 \pi \kappa \sqrt{\frac{j 2 \pi f}{D} + h^2} )^{-1}$ in standard models \cite{quadrupoles,cahillmodel} \revision{that account for the heat diffusion in the metal transducer, thermal contact resistivity $r_{\text{ms}}$ of the metal/semiconductor interface and Gaussian shape of the laser beams} for analysis of the sample structures employed in pulsed laser experiments.
\par
We have validated the stability and accuracy of the various numerical operations employed in the formalism by running our truncated L\'evy simulator in quasi-diffusive regime (setting $\alpha = 1.999$ and $u_{\text{BD}} = 10\,$nm) for a semi-infinte InGaAs substrate with 50$\,$nm Al transducer. Both magnitude and phase of the numerically obtained single pulse responses at the transducer top surface stayed within $\pm$0.3\% of analytical Fourier solutions anywhere over the 100$\,$kHz--1$\,$THz frequency band.
\subsection*{Single pulse response characteristics}
As could be expected, pure L\'evy behavior dominates truncated L\'evy flights at early times $t \ll t_{\text{BD}}$. One-dimensional single pulse responses exhibit an elevated energy density at the heat source that drops as $t^{-1/\alpha}$ (Fig 1b). This corresponds to the quasi-ballistic regime in the BTE solutions, which exhibit the same $t^{-1/\alpha}$ trend at the heat source and superdiffusive thermal energy displacement $\sigma^2(t) \sim t^{3-\alpha}$ inside the medium. Both anomalous fractional time exponents signal a complex interplay between ballistic jumps and random scatterings. This offers a perspective that is quite different from recent literature. In some experimental configurations, quasi-ballistic transport can be interpreted as a lack of scattering at short distances followed by regular Fourier diffusion from a heat source whose effective dimension is inflated by the dominant phonon MFP \cite{siemens}. For the problem studied here this is clearly not the case: the transport dynamics gradually evolve in space and time from fractal L\'evy superdiffusion to regular Fourier diffusion. The transition is virtually complete at $t = t_{\text{BD}}$ (Fig. 1b). Similar evolutions are visible in 3D heat flow configurations (Fig. 1c).
\par
Quasi-ballistic effects can be mostly attributed to phonon modes whose mean free paths extend beyond the charactersistic dimension of the thermal gradient \cite{cahill,MFPspectroscopy,malen}. We therefore expect $u_{\text{BD}}$ to be situated somewhere between the median and upper regions of the MFP spectrum. The fractal dimension $\alpha$, on the other hand, is directly associated with the order $n$ of the dominant phonon scattering mechanism $\tau \sim \omega^{-n}$ as demonstrated by our BTE analysis \cite{part1}.
\par
Figure 1d shows the calculated 3D impulse resonse inside semi-infinite InGaAs. The temperature at and near the heat source is significantly larger than Fourier predictions while at intermediate depths in the material the truncated L\'evy thermal field is substantially smaller than the diffusive counterpart. Deviations persist over 1$\,\mu$s time and 10$\,\mu$m length scales and can therefore have interesting implications for the thermal performance and monitoring of nanoscale devices. Details will be investigated elsewhere. The formalism provides valuable predictions about how the L\'evy dynamics of the quasi-ballistic transport influence the internal heat flow inside the medium. Currently, there are no measurement results available with sufficient resolution to enable direct comparison with Fig. 1d, but these and related effects may be further verified with future experiments.
\subsection*{Results}
\myfigwide[!htb]{width=0.8\textwidth}{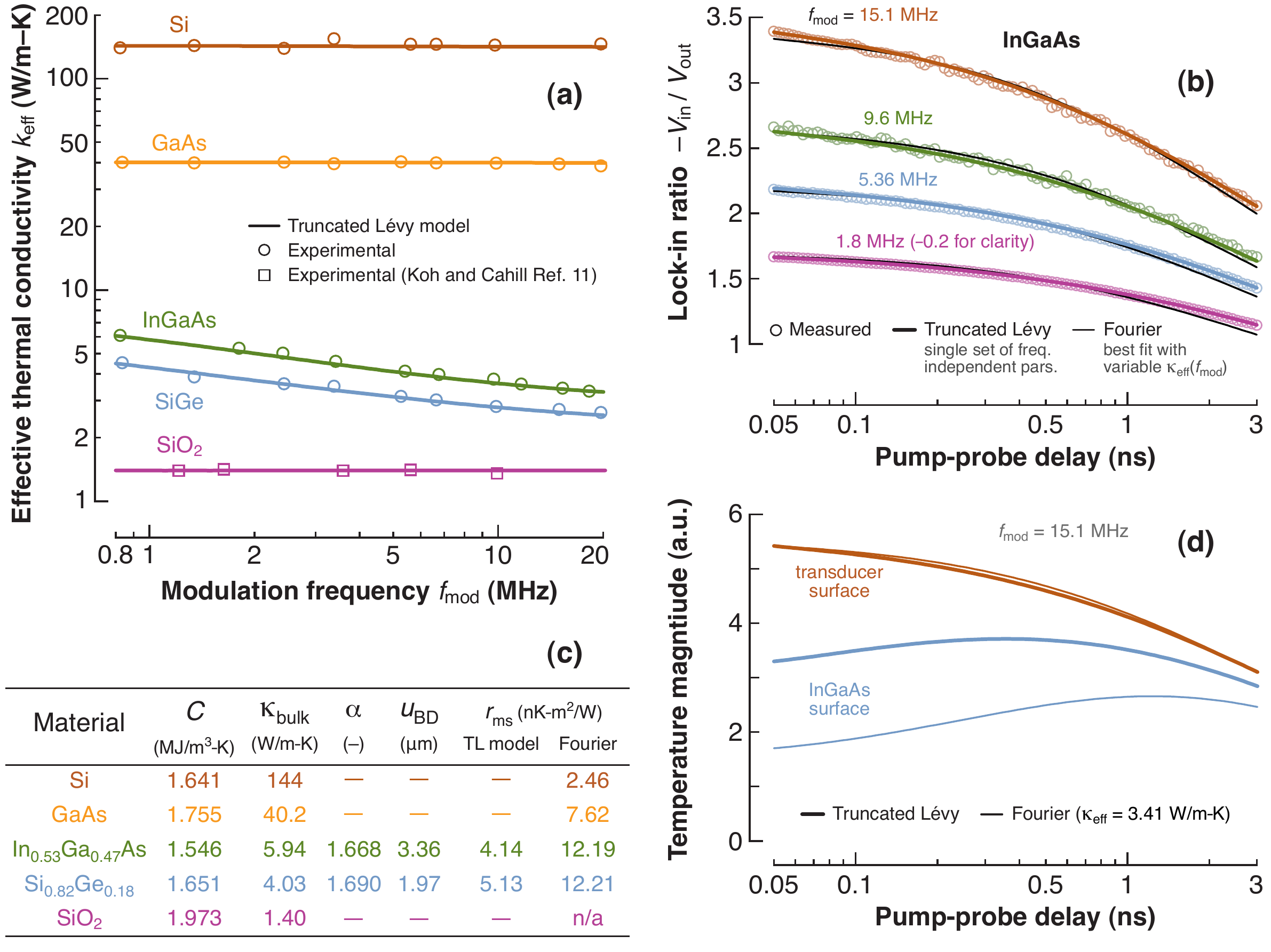}{\textbf{Time domain thermoreflectance characterization of various thin films at room temperature.} (a), Apparent modulation frequency dependence of thermal conductivity. (b), Experimental and theoretical raw transient data for InGaAs at four modulation frequencies. (c), Material heat capacities and identified model parameters. Entries marked `---' signify diffusive regime (the raw measurement data can be fitted by a Fourier model with constant conductivity). (d), Modified Fourier theory, i.e. a regular diffusive model with adjusted thermal conductivity, is unable to properly describe the quasi-ballistic regime. While such an approach may appear adequate based on fittings of the transducer response, it does not give a good representation of the intricate dynamics at the semiconductor surface. Similar trends are observed at all modulation frequencies.}%
We apply our formalism to TDTR observations of quasi-ballistic effects in semiconductor alloys. This measurement technique employs modulated femtosecond laser pulse trains to perform thermal characterization \cite{cahill,cahillmodel}. A pump beam, modulated at frequency $f_{\text{mod}}$, heats a metal transducer deposited on the semiconductor sample. Lock-in detection at $f_{\text{mod}}$ of the reflected probe beam records the thermal decay of the transducer surface. A mechanical delay line allows to vary the relative arrival times of the pump and probe pulses at the sample surface with picosecond resolution. Additional details and a schematic drawing of our measurement system are available elsewhere \cite{PCS}. Standard mathematical manipulations of the single pulse response \cite{cahillmodel,heterodyne} provide theoretical model expressions for the in-phase $V_{\text{in}}(t_{\text{pp}})$ and out-of-phase $V_{\text{out}}(t_{\text{pp}})$ lock-in signal components at a given modulation frequency as a function of pump-probe delay $t_{\text{pp}}$. These are then fitted to the measured counterparts to identify the thermal properties of the sample. The actual identification process is typically performed on the ratio $-V_{\text{in}}/V_{\text{out}}$. This acts as signal normalization and reduces the influence of experimental artifacts \cite{cahill}. Conventional Fourier analysis first extracts the thermal resistivity $r_{\text{ms}}$ of the metal-semiconductor interface from data at high modulation frequency, where the sensitivity to $r_{\text{ms}}$ is highest, and then identifies effective themal conductivities $k_{\text{eff}}(f_{\text{mod}})$ \cite{cahill}. Our truncated L\'evy approach collectively identifies two quasi-ballistic parameters ($\alpha$, $u_{\text{BD}}$) and bulk thermal properties ($\kappa_{\text{bulk}}$, $r_{\text{ms}}$) by minimizing the cumulative fitting error between measured and theoretical ratio curves over full pump-probe delay and all modulation frequencies. Results of the TDTR analysis of several semiconductor films at room temperature are presented in Fig. 2.
\par
The effective conductivity of SiGe and InGaAs drops strongly with modulation frequency (Fig. 2a), consistent with earlier reports \cite{cahill}. Single crystal semiconductors such as Si and GaAs and amorphous materials like SiO$_2$, by contrast, exhibit an essentially constant conductivity. The distinction can be understood by noting that in alloys, Rayleigh scattering increases the relative importance of long wavelength phonons \cite{cahill}. Theoretical calculations indicate the median MFP to be $\approx 5\,\mu$m in SiGe \cite{MFP1} versus $\approx 500\,$nm in Si and GaAs \cite{MFP1,MFP2}. The combination of low diffusivity and long MFPs in alloys facilitates overlap of the dominant experimental thermal penetration depth $\ell = \sqrt{D/(\pi f_{\text{mod}})}$ with the phonon spectrum over the achievable $f_{\text{mod}}$ range, inducing notable quasi-ballistic effects.
\par
Crucially, the truncated L\'evy model provides an accurate match with raw measurement data (Fig. 2b), and outperforms the best fitting modified Fourier approaches in which both $\kappa_{\text{eff}}$ and $r_{\text{ms,eff}}$ are suitably varied with modulation frequency (Fig. 3). \revision{Note that the improved fitting performance is achieved with fewer fitting parameters. The TL model collectively fits the raw measurement data across all modulation frequencies with just 4 numbers ($\alpha, u_{\text{BD}}, \kappa_{\text{bulk}}, r_{\text{ms}}$) while a modified Fourier analysis requires 2 parameters ($\kappa_{\text{eff}},r_{\text{ms,eff}}$) \textit{per modulation frequency}. Our experiments on InGaAs, for example, consist of 7 ratio curves. The best fitting Fourier interpretation (in which both effective conductivity and interface resistance must drop by over 40\% from lowest to highest modulation frequency) requires 14 parameters to describe the data yet provides an inferior fit compared to TL at every single frequency.} We validated the L\'evy parameter extraction, listed in Fig. 2c, by verifying that the cumulative fitting error attains a global minimum for a well defined $(\alpha,u_{\text{BD}})$ combination. The experimentally identified fractal dimensions, $\alpha = 1.67$ for InGaAs and $\alpha = 1.69$ for SiGe, are in near perfect agreement with ab initio BTE predictions (1.67 and 1.65 respectively). Meanwhile, $u_{\text{BD}}$ values around a few microns are found, on the order of the median MFPs.
\myfig[!htb]{width=0.4\textwidth}{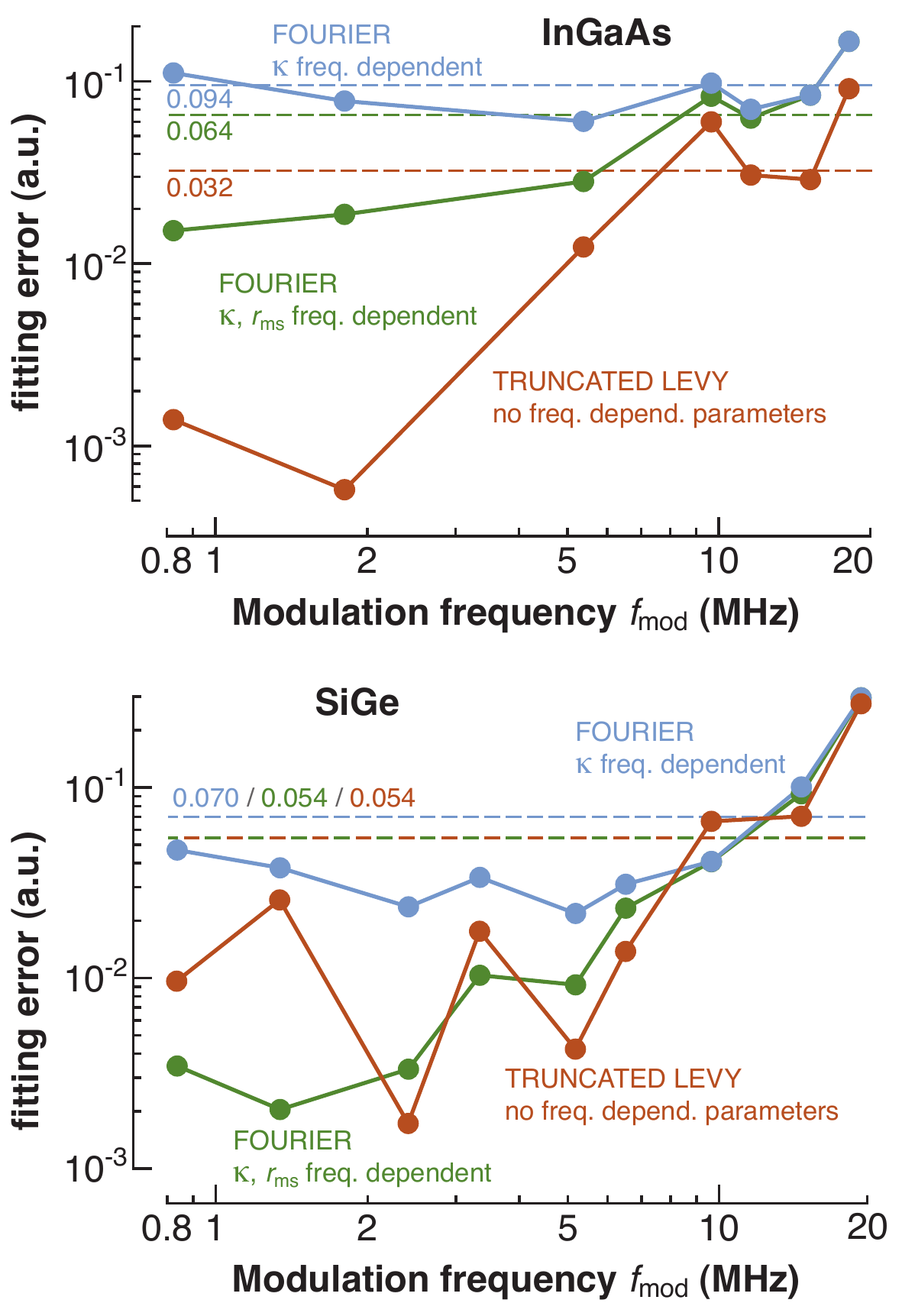}{Raw measurement data fitting performance over full pump-probe delay for several theoretical models. The dashed lines indicate the overall average fitting error across all modulation frequencies.}
\par
Conventional analysis with modified Fourier theory leaves some discrepancies at short and long pump-probe delays (Fig. 2b) but achieves an otherwise reasonable fit of the thermal response at the transducer surface. This has lead to the notion that this approach provides an adequate characterization of the quasi-ballistic transport. Recent analysis of thermal transient grating experiments has shown that most of such measurements probe a weakly quasi-ballistic regime in which the use of modified Fourier theory is formally justified by the BTE \cite{minnichBTE}. In thermoreflectance configurations considered here, however, this is not the case. Fourier theory with adjusted conductivity still maintains the Gaussian shape of the energy density and underlying Brownian energy motion, which inherently differ from L\'evy superdiffusion. As a result, this approach provides a poor representation of the quasi-ballistic dynamics at the semiconductor surface (Fig. 2d).
\subsection*{Probing fractal L\'evy dynamics}
The superior fitting performance observed above indicates that the truncated L\'evy formalism incorporates all essential physics required to gain a good understanding of non-diffusive heat flow in thermoreflectance experiments. In addition, the obtained results offer experimental validation for the fractal L\'evy nature of the quasi-ballistic thermal dynamics in semiconductor alloys, as theoretically predicted by our BTE analysis.
\par
Interestingly, the frequency dependence of effective thermal conductivity as observed by conventional Fourier analysis offers another direct manifestation of L\'evy superdiffusion. Simple 1D relations suggest that $\kappa_{\text{eff}} \sim D_{\alpha}^{{2}/{\alpha}} f_{\text{mod}}^{1 - 2/\alpha}$, and the experimental results shown in Fig. 2a can indeed be fitted quite well with a power law \cite{part1}. The resulting $\alpha$ and $D_{\alpha}$ values, together with those obtained independently through full 3D TL fitting of the raw measurement data and 1D ab initio BTE modeling, are summarized in Table \ref{tabalpha}. The good agreement between the values is testament to the capabilities of the TL formalism, and demonstrates that the presence of fractal L\'evy superdiffusion offers a consistent explanation of quasi-ballistic heat effects in semiconductor alloys.
\begin{table*}[!htb]
\caption{Identified fractal properties of quasi-ballistic thermal transport in semiconductor alloys.} \label{tabalpha}
\begin{tabular}{ccccc}
\\
\hline
& \multicolumn{2}{c}{\textbf{In$_{\text{0.53}}$Ga$_{\text{0.47}}$As}} & \multicolumn{2}{c}{\textbf{Si$_{\text{0.82}}$Ge$_{\text{0.18}}$}} \\
& $\alpha$ & $D_{\alpha}$ [$\times$10$^{-4}$m$^{\alpha}$/s] & $\alpha$ & $D_{\alpha}$ [$\times$10$^{-4}$m$^{\alpha}$/s]\\
\hline
\textbf{3D TL fit of raw meas. data$^{\ast}$} & 1.67 & 3.92 & 1.69 & 2.17\\
\textbf{power law fit of $\kappa_{\mathrm{eff}}(f_{\mathrm{mod}})$ \cite{part1}} & 1.67 & 5.98 & 1.71 & 2.17\\
\textbf{ab initio 1D BTE prediction \cite{part1}} & 1.67 & 4.28 & 1.65 & 4.00\\
\hline
\end{tabular}
\\[2mm]
\footnotesize $^{\ast}$ $D_{\alpha}$ values can be identified through TL raw data fitting by inserting the measured $\alpha$, $u_{\text{BD}}$ and $D = \kappa_{\text{bulk}}/C$ into Eq. (\ref{Dalpha}).
\end{table*}
\subsection*{Further applications}
The presence of quasi-ballistic effects raises challenges for the metrology of thermal boundary (Kapitza) resistances. It is striking that truncated L\'evy identification results for $r_{\text{ms}}$ are up to three times smaller than those obtained through conventional Fourier characterization (Fig. 2c). This suggests that metal/semiconductor interfaces could be far more conductive than currently believed. Intuitively, we can argue that Fourier models will mistakenly interpret part of the quasi-ballistic heat flux suppression in the upper regions of the semiconductor as a poorer performance of the nearby metal/semiconductor interface. Closer inspection shows that the best fitting Fourier models depicted in Fig. 3 exhibit $r_{\text{ms,eff}}$ values that vary monotonically with $f_{\text{mod}}$ by almost a factor of 2. This itself hints at a problematic aspect of modified Fourier theory, as physically the thermal boundary resistance should remain constant over the considered modulation frequency range \cite{minnich,malen}. Frequency by frequency truncated L\'evy identification of the measurement data, on the other hand, produces $r_{\text{ms}}$ values that remain virtually stable. Detailed results are available elsewhere \cite{PRBinterface}. Despite seven decades of extensive research since Kapitza's pioneering work \cite{kaptiza}, comprehensive understanding of heat flow across interfaces has remained a mostly open problem \cite{interfaces}. We believe the framework presented here offers interesting potential in this context, given its ability to properly distinguish between intrinsic interface phenomena ($r_{\text{ms}}$) and adjacent quasi-ballistic effects ($\alpha$ and $u_{\text{BD}}$).
\subsection*{Conclusions}
In this work, we introduced a novel formalism for experimental analysis of quasi-ballistic heat flow. The approach describes the thermal energy motion in terms of a truncated L\'evy stochastic process. This way, the formalism captures all essential physics of the transition between quasi-ballistic and regular diffusive transport contained within the 1D BTE framework while enabling full 3D analysis of experimental observations. Excellent agreement with raw thermoreflectance data is observed, and corresponding experimental values of the fractal dimension of the L\'evy process match ab initio BTE predictions within a few percent. Our findings confirm that quasi-ballistic thermal transport in semiconductor alloys is distinctly different from conventional Brownian motion, and will lead to a better understanding of heat flow in nanoscale devices and across metal/semiconductor interfaces.
\subsection*{Acknowledgements}
We thank Sergey Denisov, Andrea Zoia, Alain Mazzolo, David Cahill, Yee Kan Koh and Vladimir Shalaev for insightful discussions. The InGaAs sample was provided by Hong Lu and Art Gossard. The SiGe sample is courtesy of Ed Croke. A.S. acknowledges funding from the Center for Energy Efficient Materials, an Energy Frontier Research Center funded by the U.S. Department of Energy, Office of Basic Energy Sciences under Award Number DE-SC0001009 which supported the work of B.V. and Y.K; DARPA/Army Research Office, Contract No. W911NF0810347 supporting the work of G.P. and A.M.; and AFOSR MURI Grant No. FA9550-08-1-0340 for partial support of experimental capabilities.


\begin{thebibliography}{34}%
\makeatletter
\providecommand \@ifxundefined [1]{%
 \@ifx{#1\undefined}
}%
\providecommand \@ifnum [1]{%
 \ifnum #1\expandafter \@firstoftwo
 \else \expandafter \@secondoftwo
 \fi
}%
\providecommand \@ifx [1]{%
 \ifx #1\expandafter \@firstoftwo
 \else \expandafter \@secondoftwo
 \fi
}%
\providecommand \natexlab [1]{#1}%
\providecommand \enquote  [1]{``#1''}%
\providecommand \bibnamefont  [1]{#1}%
\providecommand \bibfnamefont [1]{#1}%
\providecommand \citenamefont [1]{#1}%
\providecommand \href@noop [0]{\@secondoftwo}%
\providecommand \href [0]{\begingroup \@sanitize@url \@href}%
\providecommand \@href[1]{\@@startlink{#1}\@@href}%
\providecommand \@@href[1]{\endgroup#1\@@endlink}%
\providecommand \@sanitize@url [0]{\catcode `\\12\catcode `\$12\catcode
  `\&12\catcode `\#12\catcode `\^12\catcode `\_12\catcode `\%12\relax}%
\providecommand \@@startlink[1]{}%
\providecommand \@@endlink[0]{}%
\providecommand \url  [0]{\begingroup\@sanitize@url \@url }%
\providecommand \@url [1]{\endgroup\@href {#1}{\urlprefix }}%
\providecommand \urlprefix  [0]{URL }%
\providecommand \Eprint [0]{\href }%
\providecommand \doibase [0]{http://dx.doi.org/}%
\providecommand \selectlanguage [0]{\@gobble}%
\providecommand \bibinfo  [0]{\@secondoftwo}%
\providecommand \bibfield  [0]{\@secondoftwo}%
\providecommand \translation [1]{[#1]}%
\providecommand \BibitemOpen [0]{}%
\providecommand \bibitemStop [0]{}%
\providecommand \bibitemNoStop [0]{.\EOS\space}%
\providecommand \EOS [0]{\spacefactor3000\relax}%
\providecommand \BibitemShut  [1]{\csname bibitem#1\endcsname}%
\let\auto@bib@innerbib\@empty
\bibitem [{\citenamefont {Z.{J. Ziman}}(2001)}]{phonons}%
  \BibitemOpen
  \bibfield  {author} {\bibinfo {author} {\bibnamefont {Z.{J. Ziman}}},\
  }\href@noop {} {\emph {\bibinfo {title} {Electrons \& Phonons: The Theory of
  Transport Phenomena in Solids.}}}\ (\bibinfo  {publisher} {Oxford University
  Press},\ \bibinfo {address} {USA},\ \bibinfo {year} {2001})\BibitemShut
  {NoStop}%
\bibitem [{\citenamefont {Bera}\ \emph {et~al.}(2010)\citenamefont {Bera},
  \citenamefont {Mingo},\ and\ \citenamefont {Volz}}]{MFP1}%
  \BibitemOpen
  \bibfield  {author} {\bibinfo {author} {\bibfnamefont {C.}~\bibnamefont
  {Bera}}, \bibinfo {author} {\bibfnamefont {N.}~\bibnamefont {Mingo}}, \ and\
  \bibinfo {author} {\bibfnamefont {S.}~\bibnamefont {Volz}},\ }\href@noop {}
  {\bibfield  {journal} {\bibinfo  {journal} {Phys. Rev. Lett.}\ }\textbf
  {\bibinfo {volume} {104}},\ \bibinfo {pages} {15502} (\bibinfo {year}
  {2010})}\BibitemShut {NoStop}%
\bibitem [{\citenamefont {Luo}\ and\ \citenamefont {Chen}(2013)}]{MFP2}%
  \BibitemOpen
  \bibfield  {author} {\bibinfo {author} {\bibfnamefont {T.}~\bibnamefont
  {Luo}}\ and\ \bibinfo {author} {\bibfnamefont {G.}~\bibnamefont {Chen}},\
  }\href@noop {} {\bibfield  {journal} {\bibinfo  {journal} {Phys. Chem. Chem.
  Phys.}\ }\textbf {\bibinfo {volume} {15}},\ \bibinfo {pages} {3389} (\bibinfo
  {year} {2013})}\BibitemShut {NoStop}%
\bibitem [{\citenamefont {Pop}\ \emph {et~al.}(2006)\citenamefont {Pop},
  \citenamefont {Sinha},\ and\ \citenamefont {K.{E. Goodson}}}]{devices}%
  \BibitemOpen
  \bibfield  {author} {\bibinfo {author} {\bibfnamefont {E.}~\bibnamefont
  {Pop}}, \bibinfo {author} {\bibfnamefont {S.}~\bibnamefont {Sinha}}, \ and\
  \bibinfo {author} {\bibnamefont {K.{E. Goodson}}},\ }\href@noop {} {\bibfield
   {journal} {\bibinfo  {journal} {Proc. IEEE}\ }\textbf {\bibinfo {volume}
  {94}},\ \bibinfo {pages} {1587} (\bibinfo {year} {2006})}\BibitemShut
  {NoStop}%
\bibitem [{\citenamefont {A.{J. Minnich}}\ \emph
  {et~al.}(2011{\natexlab{a}})\citenamefont {A.{J. Minnich}}, \citenamefont
  {Chen}, \citenamefont {Mansoor},\ and\ \citenamefont {B.{S.
  Yilbas}}}]{boltzmann}%
  \BibitemOpen
  \bibfield  {author} {\bibinfo {author} {\bibnamefont {A.{J. Minnich}}},
  \bibinfo {author} {\bibfnamefont {G.}~\bibnamefont {Chen}}, \bibinfo {author}
  {\bibfnamefont {S.}~\bibnamefont {Mansoor}}, \ and\ \bibinfo {author}
  {\bibnamefont {B.{S. Yilbas}}},\ }\href@noop {} {\bibfield  {journal}
  {\bibinfo  {journal} {Phys. Rev. B}\ }\textbf {\bibinfo {volume} {84}},\
  \bibinfo {pages} {235207} (\bibinfo {year} {2011}{\natexlab{a}})}\BibitemShut
  {NoStop}%
\bibitem [{\citenamefont {D.{D. Joseph}}\ and\ \citenamefont
  {Preziosi}(1989)}]{heatwaves}%
  \BibitemOpen
  \bibfield  {author} {\bibinfo {author} {\bibnamefont {D.{D. Joseph}}}\ and\
  \bibinfo {author} {\bibfnamefont {L.}~\bibnamefont {Preziosi}},\ }\href@noop
  {} {\bibfield  {journal} {\bibinfo  {journal} {Rev. Mod. Phys.}\ }\textbf
  {\bibinfo {volume} {61}},\ \bibinfo {pages} {41} (\bibinfo {year}
  {1989})}\BibitemShut {NoStop}%
\bibitem [{\citenamefont {Chen}(2001)}]{ballisticdiffusive}%
  \BibitemOpen
  \bibfield  {author} {\bibinfo {author} {\bibfnamefont {G.}~\bibnamefont
  {Chen}},\ }\href@noop {} {\bibfield  {journal} {\bibinfo  {journal} {Phys.
  Rev. Lett.}\ }\textbf {\bibinfo {volume} {86}},\ \bibinfo {pages} {2297}
  (\bibinfo {year} {2001})}\BibitemShut {NoStop}%
\bibitem [{\citenamefont {M.{E. Siemens}}\ \emph {et~al.}(2010)\citenamefont
  {M.{E. Siemens}}, \citenamefont {Li}, \citenamefont {Yang}, \citenamefont
  {K.{A. Nelson}}, \citenamefont {E.{H. Anderson}}, \citenamefont {M.{M.
  Murnane}},\ and\ \citenamefont {{C. Kapteyn}}}]{siemens}%
  \BibitemOpen
  \bibfield  {author} {\bibinfo {author} {\bibnamefont {M.{E. Siemens}}},
  \bibinfo {author} {\bibfnamefont {Q.}~\bibnamefont {Li}}, \bibinfo {author}
  {\bibfnamefont {R.}~\bibnamefont {Yang}}, \bibinfo {author} {\bibnamefont
  {K.{A. Nelson}}}, \bibinfo {author} {\bibnamefont {E.{H. Anderson}}},
  \bibinfo {author} {\bibnamefont {M.{M. Murnane}}}, \ and\ \bibinfo {author}
  {\bibfnamefont {H.}~\bibnamefont {{C. Kapteyn}}},\ }\href@noop {} {\bibfield
  {journal} {\bibinfo  {journal} {Nature Mater.}\ }\textbf {\bibinfo {volume}
  {9}},\ \bibinfo {pages} {26} (\bibinfo {year} {2010})}\BibitemShut {NoStop}%
\bibitem [{\citenamefont {A.{J. Minnich}}\ \emph
  {et~al.}(2011{\natexlab{b}})\citenamefont {A.{J. Minnich}}, \citenamefont
  {J.{A. Johnson}}, \citenamefont {A.{J. Schmidt}}, \citenamefont {Esfarjani},
  \citenamefont {M.{S. Dresselhaus}},\ and\ \citenamefont {Chen}}]{minnich}%
  \BibitemOpen
  \bibfield  {author} {\bibinfo {author} {\bibnamefont {A.{J. Minnich}}},
  \bibinfo {author} {\bibnamefont {J.{A. Johnson}}}, \bibinfo {author}
  {\bibnamefont {A.{J. Schmidt}}}, \bibinfo {author} {\bibfnamefont
  {K.}~\bibnamefont {Esfarjani}}, \bibinfo {author} {\bibfnamefont
  {K.}~\bibnamefont {M.{S. Dresselhaus}}}, \ and\ \bibinfo {author}
  {\bibfnamefont {G.}~\bibnamefont {Chen}},\ }\href@noop {} {\bibfield
  {journal} {\bibinfo  {journal} {Phys. Rev. Lett.}\ }\textbf {\bibinfo
  {volume} {107}},\ \bibinfo {pages} {095901} (\bibinfo {year}
  {2011}{\natexlab{b}})}\BibitemShut {NoStop}%
\bibitem [{\citenamefont {J.{A. Johnson}}\ \emph {et~al.}(2013)\citenamefont
  {J.{A. Johnson}}, \citenamefont {J.{A. Johnson}}, \citenamefont {A.{A.
  Maznev}}, \citenamefont {Cuffe}, \citenamefont {J.{K. Eliason}},
  \citenamefont {A.{J. Minnich}}, \citenamefont {Kehoe}, \citenamefont {C.{M.
  Sotomayor Torres}}, \citenamefont {Chen},\ and\ \citenamefont {K.{A.
  Nelson}}}]{johnson}%
  \BibitemOpen
  \bibfield  {author} {\bibinfo {author} {\bibnamefont {J.{A. Johnson}}},
  \bibinfo {author} {\bibnamefont {J.{A. Johnson}}}, \bibinfo {author}
  {\bibnamefont {A.{A. Maznev}}}, \bibinfo {author} {\bibfnamefont
  {J.}~\bibnamefont {Cuffe}}, \bibinfo {author} {\bibnamefont {J.{K.
  Eliason}}}, \bibinfo {author} {\bibnamefont {A.{J. Minnich}}}, \bibinfo
  {author} {\bibfnamefont {T.}~\bibnamefont {Kehoe}}, \bibinfo {author}
  {\bibnamefont {C.{M. Sotomayor Torres}}}, \bibinfo {author} {\bibfnamefont
  {G.}~\bibnamefont {Chen}}, \ and\ \bibinfo {author} {\bibnamefont {K.{A.
  Nelson}}},\ }\href@noop {} {\bibfield  {journal} {\bibinfo  {journal} {Phys.
  Rev. Lett.}\ }\textbf {\bibinfo {volume} {110}},\ \bibinfo {pages} {025901}
  (\bibinfo {year} {2013})}\BibitemShut {NoStop}%
\bibitem [{\citenamefont {Y.{K. Koh}}\ and\ \citenamefont {D.{G.
  Cahill}}(2007)}]{cahill}%
  \BibitemOpen
  \bibfield  {author} {\bibinfo {author} {\bibnamefont {Y.{K. Koh}}}\ and\
  \bibinfo {author} {\bibnamefont {D.{G. Cahill}}},\ }\href@noop {} {\bibfield
  {journal} {\bibinfo  {journal} {Phys. Rev. B}\ }\textbf {\bibinfo {volume}
  {76}},\ \bibinfo {pages} {075207} (\bibinfo {year} {2007})}\BibitemShut
  {NoStop}%
\bibitem [{\citenamefont {K.{T. Regner}}\ \emph {et~al.}(2013)\citenamefont
  {K.{T. Regner}}, \citenamefont {D.{P. Sellan}}, \citenamefont {Su},
  \citenamefont {C.{H. Amon}}, \citenamefont {A.{J.H. McGaughey}},\ and\
  \citenamefont {J.{A. Malen}}}]{malen}%
  \BibitemOpen
  \bibfield  {author} {\bibinfo {author} {\bibnamefont {K.{T. Regner}}},
  \bibinfo {author} {\bibnamefont {D.{P. Sellan}}}, \bibinfo {author}
  {\bibfnamefont {Z.}~\bibnamefont {Su}}, \bibinfo {author} {\bibnamefont
  {C.{H. Amon}}}, \bibinfo {author} {\bibnamefont {A.{J.H. McGaughey}}}, \ and\
  \bibinfo {author} {\bibnamefont {J.{A. Malen}}},\ }\href@noop {} {\bibfield
  {journal} {\bibinfo  {journal} {Nat. Commun.}\ }\textbf {\bibinfo {volume}
  {4}},\ \bibinfo {pages} {1640} (\bibinfo {year} {2013})}\BibitemShut
  {NoStop}%
\bibitem [{\citenamefont {Vermeersch}\ \emph {et~al.}()\citenamefont
  {Vermeersch}, \citenamefont {Carrete}, \citenamefont {Mingo},\ and\
  \citenamefont {Shakouri}}]{part1}%
  \BibitemOpen
  \bibfield  {author} {\bibinfo {author} {\bibfnamefont {B.}~\bibnamefont
  {Vermeersch}}, \bibinfo {author} {\bibfnamefont {J.}~\bibnamefont {Carrete}},
  \bibinfo {author} {\bibfnamefont {N.}~\bibnamefont {Mingo}}, \ and\ \bibinfo
  {author} {\bibfnamefont {A.}~\bibnamefont {Shakouri}},\ }\href@noop {}
  {\bibinfo  {journal} {Superdiffusive heat conduction in semiconductor alloys
  -- I. Theoretical foundations. Submitted to Phys. Rev. B. Preprint available
  online at arXiv:1406.7341}\ }\BibitemShut {NoStop}%
\bibitem [{\citenamefont {Klafter}\ \emph {et~al.}(1987)\citenamefont
  {Klafter}, \citenamefont {Blumen},\ and\ \citenamefont {M.{F.
  Shlesinger}}}]{CTRWmaster}%
  \BibitemOpen
\bibfield  {journal} {  }\bibfield  {author} {\bibinfo {author} {\bibfnamefont
  {J.}~\bibnamefont {Klafter}}, \bibinfo {author} {\bibfnamefont
  {A.}~\bibnamefont {Blumen}}, \ and\ \bibinfo {author} {\bibnamefont {M.{F.
  Shlesinger}}},\ }\href@noop {} {\bibfield  {journal} {\bibinfo  {journal}
  {Phys. Rev. A}\ }\textbf {\bibinfo {volume} {35}},\ \bibinfo {pages} {3081}
  (\bibinfo {year} {1987})}\BibitemShut {NoStop}%
\bibitem [{\citenamefont {M.{F. Shlesinger}}\ \emph {et~al.}(1999)\citenamefont
  {M.{F. Shlesinger}}, \citenamefont {Klafter},\ and\ \citenamefont
  {Zumofen}}]{brownian}%
  \BibitemOpen
  \bibfield  {author} {\bibinfo {author} {\bibnamefont {M.{F. Shlesinger}}},
  \bibinfo {author} {\bibfnamefont {J.}~\bibnamefont {Klafter}}, \ and\
  \bibinfo {author} {\bibfnamefont {G.}~\bibnamefont {Zumofen}},\ }\href@noop
  {} {\bibfield  {journal} {\bibinfo  {journal} {Am. J. Phys.}\ }\textbf
  {\bibinfo {volume} {67}},\ \bibinfo {pages} {1253} (\bibinfo {year}
  {1999})}\BibitemShut {NoStop}%
\bibitem [{\citenamefont {Metzler}\ and\ \citenamefont
  {Klafter}(2000)}]{fractalwalks}%
  \BibitemOpen
  \bibfield  {author} {\bibinfo {author} {\bibfnamefont {R.}~\bibnamefont
  {Metzler}}\ and\ \bibinfo {author} {\bibfnamefont {J.}~\bibnamefont
  {Klafter}},\ }\href@noop {} {\bibfield  {journal} {\bibinfo  {journal} {Phys.
  Rep.}\ }\textbf {\bibinfo {volume} {339}},\ \bibinfo {pages} {1} (\bibinfo
  {year} {2000})}\BibitemShut {NoStop}%
\bibitem [{\citenamefont {G.{M. Viswanathan}}\ \emph
  {et~al.}(1999)\citenamefont {G.{M. Viswanathan}}, \citenamefont {S.{V.
  Buldyrev}}, \citenamefont {Havlin}, \citenamefont {M.{G. E. da Luz}},
  \citenamefont {E.{P. Raposo}},\ and\ \citenamefont {H.{E.
  Stanley}}}]{animals}%
  \BibitemOpen
  \bibfield  {author} {\bibinfo {author} {\bibnamefont {G.{M. Viswanathan}}},
  \bibinfo {author} {\bibnamefont {S.{V. Buldyrev}}}, \bibinfo {author}
  {\bibfnamefont {S.}~\bibnamefont {Havlin}}, \bibinfo {author} {\bibnamefont
  {M.{G. E. da Luz}}}, \bibinfo {author} {\bibnamefont {E.{P. Raposo}}}, \ and\
  \bibinfo {author} {\bibnamefont {H.{E. Stanley}}},\ }\href@noop {} {\bibfield
   {journal} {\bibinfo  {journal} {Nature}\ }\textbf {\bibinfo {volume}
  {401}},\ \bibinfo {pages} {911} (\bibinfo {year} {1999})}\BibitemShut
  {NoStop}%
\bibitem [{\citenamefont {M.{A. Lomholt}}\ \emph {et~al.}(2005)\citenamefont
  {M.{A. Lomholt}}, \citenamefont {Ambj\"ornsson},\ and\ \citenamefont
  {Metzler}}]{DNA}%
  \BibitemOpen
  \bibfield  {author} {\bibinfo {author} {\bibnamefont {M.{A. Lomholt}}},
  \bibinfo {author} {\bibfnamefont {T.}~\bibnamefont {Ambj\"ornsson}}, \ and\
  \bibinfo {author} {\bibfnamefont {R.}~\bibnamefont {Metzler}},\ }\href@noop
  {} {\bibfield  {journal} {\bibinfo  {journal} {Phys. Rev. Lett.}\ }\textbf
  {\bibinfo {volume} {95}},\ \bibinfo {pages} {260603} (\bibinfo {year}
  {2005})}\BibitemShut {NoStop}%
\bibitem [{\citenamefont {del Castillo-Negrete}(1998)}]{fluidturbulence}%
  \BibitemOpen
  \bibfield  {author} {\bibinfo {author} {\bibfnamefont {D.}~\bibnamefont {del
  Castillo-Negrete}},\ }\href@noop {} {\bibfield  {journal} {\bibinfo
  {journal} {Phys. Fluids}\ }\textbf {\bibinfo {volume} {10}},\ \bibinfo
  {pages} {576} (\bibinfo {year} {1998})}\BibitemShut {NoStop}%
\bibitem [{\citenamefont {R.{N. Mantegna}}\ and\ \citenamefont {H.{E.
  Stanley}}(1995)}]{stockmarket}%
  \BibitemOpen
  \bibfield  {author} {\bibinfo {author} {\bibnamefont {R.{N. Mantegna}}}\ and\
  \bibinfo {author} {\bibnamefont {H.{E. Stanley}}},\ }\href@noop {} {\bibfield
   {journal} {\bibinfo  {journal} {Nature}\ }\textbf {\bibinfo {volume}
  {376}},\ \bibinfo {pages} {46} (\bibinfo {year} {1995})}\BibitemShut
  {NoStop}%
\bibitem [{\citenamefont {Dhar}\ \emph {et~al.}(2013)\citenamefont {Dhar},
  \citenamefont {Saito},\ and\ \citenamefont {Derrida}}]{atomchain}%
  \BibitemOpen
  \bibfield  {author} {\bibinfo {author} {\bibfnamefont {A.}~\bibnamefont
  {Dhar}}, \bibinfo {author} {\bibfnamefont {K.}~\bibnamefont {Saito}}, \ and\
  \bibinfo {author} {\bibfnamefont {B.}~\bibnamefont {Derrida}},\ }\href@noop
  {} {\bibfield  {journal} {\bibinfo  {journal} {Phys. Rev. E}\ }\textbf
  {\bibinfo {volume} {87}},\ \bibinfo {pages} {010103} (\bibinfo {year}
  {2013})}\BibitemShut {NoStop}%
\bibitem [{\citenamefont {R.{N. Mantegna}}\ and\ \citenamefont {H.{E.
  Stanley}}(1994)}]{truncation1}%
  \BibitemOpen
  \bibfield  {author} {\bibinfo {author} {\bibnamefont {R.{N. Mantegna}}}\ and\
  \bibinfo {author} {\bibnamefont {H.{E. Stanley}}},\ }\href@noop {} {\bibfield
   {journal} {\bibinfo  {journal} {Phys. Rev. Lett.}\ }\textbf {\bibinfo
  {volume} {73}},\ \bibinfo {pages} {2946} (\bibinfo {year}
  {1994})}\BibitemShut {NoStop}%
\bibitem [{\citenamefont {Koponen}(1995)}]{truncation2}%
  \BibitemOpen
  \bibfield  {author} {\bibinfo {author} {\bibfnamefont {I.}~\bibnamefont
  {Koponen}},\ }\href@noop {} {\bibfield  {journal} {\bibinfo  {journal} {Phys.
  Rev. E}\ }\textbf {\bibinfo {volume} {52}},\ \bibinfo {pages} {1197}
  (\bibinfo {year} {1995})}\BibitemShut {NoStop}%
\bibitem [{\citenamefont {Cartea}\ and\ \citenamefont {del
  Castillo-Negrete}(2007)}]{TLfractional}%
  \BibitemOpen
  \bibfield  {author} {\bibinfo {author} {\bibfnamefont {A.}~\bibnamefont
  {Cartea}}\ and\ \bibinfo {author} {\bibfnamefont {D.}~\bibnamefont {del
  Castillo-Negrete}},\ }\href@noop {} {\bibfield  {journal} {\bibinfo
  {journal} {Phys. Rev. E}\ }\textbf {\bibinfo {volume} {76}},\ \bibinfo
  {pages} {041105} (\bibinfo {year} {2007})}\BibitemShut {NoStop}%
\bibitem [{\citenamefont {Levin}(1997)}]{FFTcollocation}%
  \BibitemOpen
  \bibfield  {author} {\bibinfo {author} {\bibfnamefont {D.}~\bibnamefont
  {Levin}},\ }\href@noop {} {\bibfield  {journal} {\bibinfo  {journal} {Comput.
  Appl. Maths.}\ }\textbf {\bibinfo {volume} {78}},\ \bibinfo {pages} {131}
  (\bibinfo {year} {1997})}\BibitemShut {NoStop}%
\bibitem [{\citenamefont {Maillet}\ \emph {et~al.}(2000)\citenamefont
  {Maillet}, \citenamefont {Andr\'e}, \citenamefont {J.{C. Batsale}},
  \citenamefont {Degiovanni},\ and\ \citenamefont {Moyne}}]{quadrupoles}%
  \BibitemOpen
  \bibfield  {author} {\bibinfo {author} {\bibfnamefont {D.}~\bibnamefont
  {Maillet}}, \bibinfo {author} {\bibfnamefont {S.}~\bibnamefont {Andr\'e}},
  \bibinfo {author} {\bibnamefont {J.{C. Batsale}}}, \bibinfo {author}
  {\bibfnamefont {A.}~\bibnamefont {Degiovanni}}, \ and\ \bibinfo {author}
  {\bibfnamefont {C.}~\bibnamefont {Moyne}},\ }\href@noop {} {\emph {\bibinfo
  {title} {Thermal Quadrupoles: Solving the Heat Equation through Integral
  Transforms.}}}\ (\bibinfo  {publisher} {Wiley},\ \bibinfo {address}
  {Chichester},\ \bibinfo {year} {2000})\BibitemShut {NoStop}%
\bibitem [{\citenamefont {D.{G. Cahill}}(2004)}]{cahillmodel}%
  \BibitemOpen
  \bibfield  {author} {\bibinfo {author} {\bibnamefont {D.{G. Cahill}}},\
  }\href@noop {} {\bibfield  {journal} {\bibinfo  {journal} {Rev. Sci.
  Instrum.}\ }\textbf {\bibinfo {volume} {75}},\ \bibinfo {pages} {5119}
  (\bibinfo {year} {2004})}\BibitemShut {NoStop}%
\bibitem [{\citenamefont {A.{J. Minnich}}(2012)}]{MFPspectroscopy}%
  \BibitemOpen
  \bibfield  {author} {\bibinfo {author} {\bibnamefont {A.{J. Minnich}}},\
  }\href@noop {} {\bibfield  {journal} {\bibinfo  {journal} {Phys. Rev. Lett.}\
  }\textbf {\bibinfo {volume} {109}},\ \bibinfo {pages} {205901} (\bibinfo
  {year} {2012})}\BibitemShut {NoStop}%
\bibitem [{\citenamefont {Vermeersch}\ \emph {et~al.}(2013)\citenamefont
  {Vermeersch}, \citenamefont {Pernot}, \citenamefont {Lu}, \citenamefont
  {Bahk}, \citenamefont {Gossard},\ and\ \citenamefont {Shakouri}}]{PCS}%
  \BibitemOpen
  \bibfield  {author} {\bibinfo {author} {\bibfnamefont {B.}~\bibnamefont
  {Vermeersch}}, \bibinfo {author} {\bibfnamefont {G.}~\bibnamefont {Pernot}},
  \bibinfo {author} {\bibfnamefont {H.}~\bibnamefont {Lu}}, \bibinfo {author}
  {\bibfnamefont {J.-H.}\ \bibnamefont {Bahk}}, \bibinfo {author}
  {\bibfnamefont {A.}~\bibnamefont {Gossard}}, \ and\ \bibinfo {author}
  {\bibfnamefont {A.}~\bibnamefont {Shakouri}},\ }\href@noop {} {\bibfield
  {journal} {\bibinfo  {journal} {Phys. Rev. B}\ }\textbf {\bibinfo {volume}
  {88}},\ \bibinfo {pages} {214302} (\bibinfo {year} {2013})}\BibitemShut
  {NoStop}%
\bibitem [{\citenamefont {Dilhaire}\ \emph {et~al.}(2011)\citenamefont
  {Dilhaire}, \citenamefont {Pernot}, \citenamefont {Calbris}, \citenamefont
  {J.{M. Rampnoux}},\ and\ \citenamefont {Grauby}}]{heterodyne}%
  \BibitemOpen
  \bibfield  {author} {\bibinfo {author} {\bibfnamefont {S.}~\bibnamefont
  {Dilhaire}}, \bibinfo {author} {\bibfnamefont {G.}~\bibnamefont {Pernot}},
  \bibinfo {author} {\bibfnamefont {G.}~\bibnamefont {Calbris}}, \bibinfo
  {author} {\bibnamefont {J.{M. Rampnoux}}}, \ and\ \bibinfo {author}
  {\bibfnamefont {S.}~\bibnamefont {Grauby}},\ }\href@noop {} {\bibfield
  {journal} {\bibinfo  {journal} {J. Appl. Phys.}\ }\textbf {\bibinfo {volume}
  {110}},\ \bibinfo {pages} {114314} (\bibinfo {year} {2011})}\BibitemShut
  {NoStop}%
\bibitem [{\citenamefont {Hua}\ and\ \citenamefont {A.{J.
  Minnich}}(2014)}]{minnichBTE}%
  \BibitemOpen
  \bibfield  {author} {\bibinfo {author} {\bibfnamefont {C.}~\bibnamefont
  {Hua}}\ and\ \bibinfo {author} {\bibnamefont {A.{J. Minnich}}},\ }\href@noop
  {} {\bibfield  {journal} {\bibinfo  {journal} {Phys. Rev. B}\ }\textbf
  {\bibinfo {volume} {89}},\ \bibinfo {pages} {094302} (\bibinfo {year}
  {2014})}\BibitemShut {NoStop}%
\bibitem [{\citenamefont {Vermeersch}\ \emph {et~al.}(2014)\citenamefont
  {Vermeersch}, \citenamefont {A.{M.S. Mohammed}}, \citenamefont {Pernot},
  \citenamefont {Koh},\ and\ \citenamefont {Shakouri}}]{PRBinterface}%
  \BibitemOpen
  \bibfield  {author} {\bibinfo {author} {\bibfnamefont {B.}~\bibnamefont
  {Vermeersch}}, \bibinfo {author} {\bibnamefont {A.{M.S. Mohammed}}}, \bibinfo
  {author} {\bibfnamefont {G.}~\bibnamefont {Pernot}}, \bibinfo {author}
  {\bibfnamefont {Y.}~\bibnamefont {Koh}}, \ and\ \bibinfo {author}
  {\bibfnamefont {A.}~\bibnamefont {Shakouri}},\ }\href@noop {} {\bibfield
  {journal} {\bibinfo  {journal} {Phys. Rev. B}\ }\textbf {\bibinfo {volume}
  {90}},\ \bibinfo {pages} {014306} (\bibinfo {year} {2014})}\BibitemShut
  {NoStop}%
\bibitem [{\citenamefont {P.{L. Kaptiza}}(1941)}]{kaptiza}%
  \BibitemOpen
  \bibfield  {author} {\bibinfo {author} {\bibnamefont {P.{L. Kaptiza}}},\
  }\href@noop {} {\bibfield  {journal} {\bibinfo  {journal} {J. Phys.
  (Moscow)}\ }\textbf {\bibinfo {volume} {4}},\ \bibinfo {pages} {181}
  (\bibinfo {year} {1941})}\BibitemShut {NoStop}%
\bibitem [{\citenamefont {M.{D. Losego}}\ and\ \citenamefont {D.{G.
  Cahill}}(2013)}]{interfaces}%
  \BibitemOpen
  \bibfield  {author} {\bibinfo {author} {\bibnamefont {M.{D. Losego}}}\ and\
  \bibinfo {author} {\bibnamefont {D.{G. Cahill}}},\ }\href@noop {} {\bibfield
  {journal} {\bibinfo  {journal} {Nature Mater.}\ }\textbf {\bibinfo {volume}
  {12}},\ \bibinfo {pages} {382} (\bibinfo {year} {2013})}\BibitemShut
  {NoStop}%
\end{thebibliography}


%

\end{document}